\documentclass[aps,prxquantum,floatfix,twocolumn,superscriptaddress]{revtex4-2}

\usepackage{soul}
\usepackage[utf8]{inputenc}
\usepackage[T1]{fontenc}
\usepackage{lmodern}
\usepackage{latexsym}
\usepackage{graphicx}
\usepackage{rotating}
\usepackage[normalem]{ulem}
\usepackage{hyperref}
\usepackage{amsmath,amsfonts}
\usepackage{bm}
\usepackage{color}
\usepackage{xcolor}
\usepackage{float}
\usepackage{textgreek}
\usepackage{qcircuit}
\usepackage[greek,english]{babel}
\usepackage{amsmath}

\usepackage{color}

\usepackage{amsfonts}
\usepackage{mathrsfs}
\usepackage{physics}

\hypersetup{colorlinks=true,urlcolor=blue,linkcolor=blue}

\def\a{\alpha}

\def\b{\beta}
\def\D{\Delta}
\def\d{\delta}
\def\e{\epsilon}

\def\v{\nu}

\def\s{\sigma}

\def\w{\omega}

\def\srvo{SrVO$_3$}
\def\dd{\dagger}
\def\ua{\uparrow}
\def\da{\downarrow}

\providecommand{\d}[1]{\color{grey}[#1]\color{black}}

\begin{document}
\title{Anderson impurity solver integrating tensor network methods with quantum computing}

\author{Fran\c{c}ois Jamet}
\email{francois.jamet@francoisjamet.fr}
\affiliation{National  Physical  Laboratory,  Teddington,  TW11  0LW,  United  Kingdom}
\affiliation{IQM Quantum Computers France SaS, 40 Rue du Louvre, 75001 Paris, France}
\author{Lachlan P. Lindoy}
\email{lachlan.lindoy@npl.co.uk}
\affiliation{National  Physical  Laboratory,  Teddington,  TW11  0LW,  United  Kingdom}
\author{Yannic Rath}
\affiliation{National  Physical  Laboratory,  Teddington,  TW11  0LW,  United  Kingdom}
\author{Connor Lenihan}
\affiliation{National  Physical  Laboratory,  Teddington,  TW11  0LW,  United  Kingdom}
\author{Abhishek Agarwal}
\affiliation{National  Physical  Laboratory,  Teddington,  TW11  0LW,  United  Kingdom}
\author{Enrico Fontana}
\affiliation{National  Physical  Laboratory,  Teddington,  TW11  0LW,  United  Kingdom}
\affiliation{Department of Computer and Information Sciences, University of Strathclyde, 26 Richmond Street, Glasgow  G1 1XH, United Kingdom}
\author{Fedor Simkovic IV}
\affiliation{IQM Quantum Computers, Georg-Brauchle-Ring 23-25, 80992, Munich, Germany}
\author{Baptiste Anselme Martin}
\affiliation{TotalEnergies, Tour Coupole - 2 place Jean Millier 92078 Paris la Défense, France}
\affiliation{Universite Paris-Saclay, CNRS, Laboratoire de Physique des Solides, 91405, Orsay, France}
\author{Ivan Rungger}
\email{ivan.rungger@npl.co.uk}
\affiliation{National  Physical  Laboratory,  Teddington,  TW11  0LW,  United  Kingdom}

\begin{abstract}
  Solving the Anderson impurity model typically involves a two-step process, where one first calculates the ground state of the Hamiltonian, and then computes its dynamical properties to obtain the Green's function. Here we propose a hybrid classical/quantum algorithm where the first step is performed using a classical computer to obtain the tensor network ground state as well as its quantum circuit representation, and the second step is executed on the quantum computer to obtain the Green's function. Our algorithm exploits the efficiency of tensor networks for preparing ground states on classical computers, and takes advantage of quantum processors for the evaluation of the time evolution, which can become intractable on classical computers.
  We demonstrate the algorithm using 24 qubits on a quantum computing emulator for \srvo~with a multi-orbital Anderson impurity model within the dynamical mean field theory. The tensor network based ground state quantum circuit preparation algorithm can also be performed for up to 60 qubits with our available computing resources, while the state vector emulation of the quantum algorithm for time evolution is beyond what is accessible with such resources.
  We show that, provided the tensor network calculation is able to accurately obtain the ground state energy, this scheme does not require a perfect reproduction of the ground state wave function on the quantum circuit to give an accurate Green's function. This hybrid approach may lead to quantum advantage in materials simulations where the ground state can be computed classically, but where the dynamical properties cannot.
\end{abstract}

\maketitle

\section{Introduction}
The simulation of strongly correlated quantum systems is one of the most promising applications of quantum computers~\cite{Feynman1982,lloyd_quantum_simulator}. This is due to the potential for quantum processors to perform computations using exponentially fewer resources than classical computers for such systems. The most widely used classical computing method for simulations of real materials with strongly correlated electrons is based on the dynamical mean field theory (DMFT), which self-consistently maps the periodic Hubbard model onto an effective Anderson impurity model (AIM) containing only a few interacting impurity sites embedded in a non-interacting bath~\cite{dmft,dmft4,dmft3,dmft2}. Despite the development of a large number of impurity solvers over the years~\cite{ctqmc,FTN1,ttn1,PhysRevX.5.041032,mps_cheb1,mps_cheb2,mps_cheb3,dmft_impurity_solver_ed,dmft_impurity_solver_nrg1,dmft_impurity_solver_nrg2, dmft_impurity_solver_qmc1}, none of them have achieved universal applicability. Therefore, the solution of the AIM remains a challenging problem for classical computers, where a quantum computer may have an advantage. Hybrid quantum-classical approaches are the most promising way to achieve such advantage, where only those parts of the algorithm which are the most challenging for a classical computer, are executed on a quantum computer~\cite{troyer,keen2022hybrid}.
One successful class of classical computing impurity solvers includes algorithms which make use of tensor network (TN) methods~\cite{FTN1,ttn1,PhysRevX.5.041032,mps_cheb1,mps_cheb2,mps_cheb3}, which are efficient in obtaining the ground state (GS) and also the first few excited states of the system. However, the time evolution of complex impurity models required as part of the AIM solver remains a challenge due to the rapid growth of the bond dimension with time, making long-time calculations unfeasible~\cite{PhysRevE.71.036102,PhysRevLett.97.150404}.

In this article we propose an approach that combines classical and quantum computing resources to address the problem of multi-orbital AIMs. The proposed algorithm entails obtaining the GS of the multi-orbital AIM using a TN representation of the wave function~\cite{DMRG_spectral,FTN1,ttn1} on a conventional processor, and subsequently a quantum circuit is constructed to represent this wave function. The form of this quantum circuit is also constructed using a conventional processor. Finally, one uses a quantum computer to simulate the time evolution of the prepared initial quantum states.  In doing so, we are able to compute dynamical observables such as the Green's function (GF), a central mathematical object in materials calculations.
Several physical observables can be computed from the GF, such as the density of states (DOS)~\cite{PhysRevB.85.20513}.
This algorithm has the potential to surpass existing impurity solvers in situations where TN-based techniques can compute the GS, but where the growth in entanglement entropy observed during the course of time evolution~\cite{ttn1}, and the subsequent increase in bond dimension~\cite{mps_cheb2}, limits the timescale over which dynamics can be obtained~\cite{mps_cheb3}. 

\begin{figure*}
  \includegraphics[width=\textwidth]{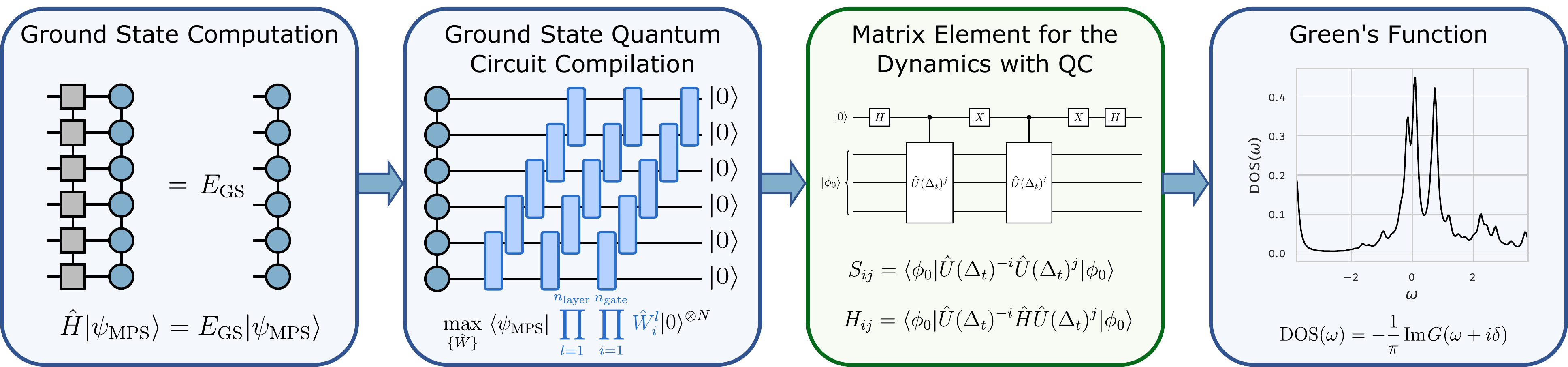}
  \caption{
    Schematic of the combined classical/quantum algorithm for computing the Green's function of the Anderson impurity model. Blue panels indicate the parts of the algorithm that are run on the classical computer, while the green panel indicates the steps to be executed on a quantum computer. The algorithm consists of four sequential steps (shown as separate panels):
    1. the computation of the ground state energy and MPS representation of the wave function using the DMRG algorithm (classical);
    2. the construction of a parametric quantum circuit representation of this wave function (classical);
    3. the simulation of the dynamics by evaluation of the matrix elements through time evolution (quantum);
    4. the evaluation of a continued fraction expansion of the Green's function using the computed matrix elements (classical).
  }
  \label{fig:schematic}
\end{figure*}

\section{Hybrid-algorithm for the calculation of Green's functions}
The Hamiltonian of the AIM is given by~\cite{dmft,dmft4,PhysRevB.51.12880}
\begin{align}
  H = & \sum^{n_{\text{imp}}}_{ij}\sum_{\s} \e_{ij} \hat{c}^\dd_{i\s} \hat{c}_{j\s} +\sum^{n_{\text{imp}}}_{i} U n_{i \ua} n_{i \da} \nonumber \\
              &+\sum^{n_{\text{imp}}}_{i > j, } \sum_{\s}(U-2J) n_{i \s} n_{j \bar{\s}} + (U-3J) n_{i \s} n_{j \s}\nonumber \\
              &-\sum^{n_{\text{imp}}}_{i > j} \sum_{\s} J c^\dagger_{i \s} c_{i \bar{\s}} c^\dagger_{j \bar{\s}} c_{j \s}  - J c^\dagger_{i \s} c^\dagger_{i \bar{\s}} c_{j \s}  c_{j \bar{\s}}.  \nonumber\\
 +& \sum^{n_{\text{imp}}}_{i}\sum^{n_b}_{j}\sum_\s V_{ij} (\hat{c}^\dd_{i\s} \hat{d}_{j\s}+ \hat{d}_{j\s}^\dd \hat{c}_{i\s}) +  \sum^{n_b}_{ij}\sum_\s \e^d_{ij} \hat{d}^\dd_{i\s}\hat{d}_{j\s}\, ,
  \label{eq:H}
\end{align}

where $\hat{c}_{i\sigma}^\dd$, $\hat{c}_{i\sigma}$ ($\hat{d}^{\dd}_{i\sigma}$, $\hat{d}_{i\sigma}$) are the creation and annihilation operators of an electron of spin $\sigma$ on the $i$-th impurity site (bath site); $\e_{ij}$ is the hopping between the impurity orbitals, $U$ and $J$ are the parameters of the Hubbard-Kanamori interaction between the orbitals, $\e^{d}_{ij}$ is the hopping matrix between the bath orbitals, and $V_{ij}$ is the hybridization matrix between impurity orbitals and bath orbitals. Differently from a general chemistry Hamiltonian, the interaction term in the AIM is confined to the impurity sites, thereby considerably reducing the number of terms in the Hamiltonian, which is proportional to $n_{\text{imp}}^2 {+} (n_{\text{imp}}{+}n_{\text{bath}})^2$, where $n_{\text{imp}}$ ($n_{\text{bath}}$) is the number of impurity sites (bath sites).

The zero temperature greater and lesser GFs are defined as
\begin{align}
  G^>_{\a\b}(z) & = \bra{\text{GS}}\hat{c}_{\a}(z - (\hat{H} -E_{\text{GS}}))^{-1}\hat{c}^\dd_{\b}\ket{\text{GS}}, \label{eq:greater} \\
  G^<_{\a\b}(z) & = \bra{\text{GS}}\hat{c}^\dd_{\a}(z + (\hat{H} -E_{\text{GS}}))^{-1}\hat{c}_{\b}\ket{\text{GS}}, \label{eq:lesser}
\end{align}
respectively, where $\ket{\text{GS}}$ is the GS wave function, $E_{\text{GS}}$ the GS energy, and $\alpha, \beta$ indexes both spin and site indices.
These GFs can be used to obtain the retarded GF as $ G_{\a\b}(z) {=} G_{\a \b}^{>}(z) + G_{\a \b}^<(z)$. In what follows we consider only the diagonal elements of the GFs, so that $\beta=\alpha$, and for ease of notation we discard the orbital index subscript, in analogy to the approach in Ref.~\cite{qseg}, where the extension to the off-diagonal elements is also presented.
The imaginary part of the retarded GF gives the density of states (DOS), $\mathrm{DOS}(\omega)= -\frac{1}{\pi} \text{Im}G(\w + i\d)$,
for $\omega$ a real energy, and $\d$ a small positive number.
For the AIM DOS only the impurity orbitals GFs need to be computed, which significantly reduces the number of terms compared to more general models, reducing the needed computational resources.

\subsection{State Preparation}

The first step of our algorithm, schematically illustrated in the left-most panel in Fig.~\ref{fig:schematic}, is the computation of the GS wave function and energy using a TN representation. Here we use a matrix product state (MPS) representation. We note that this algorithm can be extended to other TN forms, such as tree tensor networks~\cite{ttn1} and fork tensor networks~\cite{FTN1}, which have been successfully used in the context of quantum impurity models. The MPS GS wave function of a quantum impurity model, which we denote by  $\ket{\psi_{\text{MPS}}}$, can be obtained with the density matrix renormalization group (DMRG) method~\cite{DMRG1, DMRG2, DMRG3, DMRG4}.

Once the GS is obtained with the MPS representation, we compile it into a quantum circuit to prepare the initial state on the quantum device with a series of unitary gates. This compilation to obtain the quantum circuit that allows one to prepare the state on a quantum computer is achieved entirely on classical computers, and does not require executing a variational optimization on a quantum computer.
Our state initialization scheme relies on the representation of the state as a compact tensor network, which is an efficient and scalable approach to represent quantum states on a classical computer. 

\subsubsection{Exact tensor network decomposition}
Firstly, we consider an exact decomposition of a given MPS into a ladder of many-qubitunitaries~\cite{zapata}, which may then be decomposed further into the elemental gates of the device.
This gives a representation of the prepared state $\ket{\psi_{\text{QC}}}$ as a gate sequence applied to the zero state $\ket{0}^{\otimes N}$ as
\begin{equation}
  \ket{\psi_{\text{QC}}} = \prod_{i=1}^{N-1} \hat{U}_i \ket{0}^{\otimes N},
  \label{eq:exact_MPS_compilation}
\end{equation}
where $N$ is the number of qubits.
The $N-1$ unitaries $\hat{U}_i$ can be extracted directly from the matrices of the MPS representation in its canonical form~\cite{zapata}, giving gate sequences with depth linear in the number of qubits, where the number of gates is dictated by the MPS bond dimension.
This results in a decomposition, where each unitary $\hat{U}_i$ acts on a small number of $n_i$ neighbouring qubits, and the number of qubits $n_i$ is directly related to the bond dimension $\chi_i$ of the $i$-th bond in the MPS tensor representation according to $n_i = \lceil \log_2{\chi_i} \rceil + 1 $.
Having decomposed a given MPS exactly into a ladder of many-qubit, yet local, unitaries, we further compile this sequence into an elemental gate set of one- and two-qubit gates (CNOTs) by application of the Quantum Shannon Decomposition (QSD)~\cite{Shende_2006}.

To achieve the preparation of an approximation of the target state with lower circuit gate count we additionally employ a bond dimension truncation of the target MPS, resulting in lower resource requirements when compiling this into quantum circuit with the exact decomposition.
We truncate the MPS such that all its local bond dimensions fall below a target bond dimension, $\chi_\mathrm{max}$, which can be adjusted to achieve a required fidelity in the target state. Note that this fidelity is efficiently evaluated on the classical computer.
To truncate the target state to smaller bond dimensions, we follow standard protocols~\cite{orus2014practical}, involving the repeated application of singular values in the matrix product decomposition truncating the bond dimensions to $\chi_\mathrm{max}$ by discarding the smallest singular values.
We then maximize the fidelity between the truncated MPS and the untruncated target $\ket{\psi_{\text{MPS}}}$ by iteratively updating matrices in the truncated MPS in a single-site DMRG sweeping protocol with local updates to maximize the fidelity.

\subsubsection{Variational State Preparation}
As an alternative to the exact compilation method, which typically results in deep circuits with large gate counts, we furthermore consider an approximate compilation scheme proposed in Ref.~\cite{dov2022approximate}, and improved in Ref.~\cite{zapata}.
Ref.~\cite{dov2022approximate} introduces a protocol relying on a classical variational optimization of a circuit with layers composed of two-qubit gates in a staircase topology.
Ref.~\cite{zapata} proposes a method for iteratively adding layers of staircases to this protocol.
Similar strategies have been proposed in Refs.~\cite{PRXQuantum.2.010342} and~\cite{PhysRevResearch.4.043007}.
In this article we extend this approach to include $n_\mathrm{g}$-qubit gates, where $n_\mathrm{g}>1$ is the number of qubits on which the gate acts.
When represented as a matrix, an $n_\mathrm{g}$-qubit gate corresponds to a $2^{n_\mathrm{g}} \times 2^{n_\mathrm{g}}$ unitary matrix, which may then again be compiled down to one- and two-qubit gates using the QSD.
However, it should be noted that the circuit depth and with it the number of CNOT gates of such a decomposition increases significantly for increasing $n_\mathrm{g}$ (see Appendix~\ref{appendix:CNOT}), such that it is typically beneficial to keep $n_\mathrm{g}$ small.

The considered approach involves applying the staircase layers of $n_\mathrm{g}$-qubit gates on the initial state $\ket{0}^{\otimes N}$, defining a variational ansatz $\ket{\psi_{\text{QC}}} = \prod_{l=1}^{n_{\text{layer}}}\prod_{i=1}^{n_{\text{gate}}} \hat{W}^l_i\ket{0}^{\otimes N}$.
Here, $n_\mathrm{layer}$ is the number of staircase layers, $n_\mathrm{gate} = N + 1 - n_g$ is the number of $n_\mathrm{g}$-qubit gates per layer, and $\hat{W}^l_i$ denotes the $i$-th $n_\mathrm{g}$-qubit gate in layer $l$, which needs to be optimized to prepare the target state.

The quantum circuit is schematically illustrated with the blue-colored layers of gates in the second panel in Fig.~\ref{fig:schematic}.
The variational expressivity of the ansatz can be increased by increasing the values of $n_\mathrm{g}$ and $n_\mathrm{layer}$ to achieve the targeted accuracy in the state preparation.

The parametric circuit of staircase layers is optimized on a classical computer by maximising the fidelity, $F$, between the trial quantum state, $\ket{\psi_{\text{QC}}}$, and the MPS GS wave function, $\ket{\psi_{\text{MPS}}}$:
\begin{align}
  F[\{\hat{W}\}] & = |\braket{\psi_{\text{MPS}}}{\psi_{\text{QC}}}|^2. 
  \label{eq:loss}
\end{align}

We construct the optimized parametric circuit by considering different numbers of layers, $n_{\text{layer}}$, of $n_\mathrm{g}$-qubit gates included in the circuit.
For each ansatz, we sequentially update each $n_\mathrm{g}$-qubit gate, $\hat{W}_i^l$, to locally maximise the overlap with the MPS GS.
We optimise each $\hat{W}_i^l$ in order of increasing $i$ within a given layer, $l$, before proceeding to optimise each gate in the next layer. This sweeping process is repeated until the improvement made to the overlap with the target MPS state through a sweep falls below a set tolerance, at which point an additional staircase layer of $n_\mathrm{g}$-qubit gates is added.
To initialize a new layer, we make use of low bond dimension approximations to the MPS wavefunction~\cite{zapata}, giving a good starting point for a subsequent optimization of the circuit with the sweeping scheme.

For each of the sequential updates used in this optimization scheme, the $n_\mathrm{g}$-qubit operator $\hat{O}_i^l$ that optimizes the fidelity can be found analytically~\cite{zapata,PRXQuantum.2.010342}.
While the general analytical solution, $\hat{O}_i^l$, is not generally unitary, we can obtain a unitary approximation to maximize the fidelity by constructing the closest unitary approximation to $\hat{O}_i^l$.
This is achieved by using a singular value decomposition of $\hat{O}_i^l = U^\dd S V$, with unitary matrices $U^\dd$ and $V$, from which the optimal unitary approximation for the update of $\hat{W}_i^l$ is obtained as~\cite{dov2022approximate,zapata,Dborin2022,shirakawa2021automatic, PRXQuantum.2.010342}
\begin{equation}
  \hat{W}_i^l = U^{\dd}V.
  \label{eq:local_optimal_parameterss}
\end{equation}
Following Ref.~\cite{zapata}, we employ MPS representations of the intermediate states required to evaluate $\hat{O}_i^l$ in order to allow for efficient classical optimization of the quantum circuit.
The order of optimization used here allows to reuse the intermediate states constructed when optimizing individual tensors~\cite{zapata}, which significantly improves the efficiency of the algorithm.
The optimization process is schematically illustrated in the second panel in Fig.~\ref{fig:schematic}.
We note here that one does not use a quantum computer to perform the minimization, and that all these computations are performed using MPS representations of the quantum states.

\subsubsection{Hybrid State Preparation}
The variational state compilation requires the ability to efficiently represent intermediate states as a compact tensor network.
We found that often the application of staircase layers in the variational compilation approach leads to a growth of the bond dimension when applying the circuit in reverse to the target MPS, as is required in the variational compilation approach.
This limits the approach to the use of few staircase layers, beyond which the compilation becomes classically intractable.
This can become a limitation in the preparation of MPS for larger systems if very high fidelities are required.

As an alternative, we also consider the application of a hybrid approach, bridging between the exact decomposition and a variational preparation of the state.
By exploiting knowledge about the exact decomposition, we apply a variational compilation of intermediate states, restricting the size of the variational ansatz to small blocks, which are pieced together to compile the target.
Our approach is based on the exact decomposition of the MPS according to Eq.~\eqref{eq:exact_MPS_compilation}, where we iteratively compile the many-body unitaries $\hat{U}_i$ one after the other with a variational approach.

We define the $i$-th target state as the intermediate state obtained from the application of the first $i$ unitaries in the exact decomposition to the zero state, $\ket{\psi_{i}^\mathrm{t}} = \prod_{j=1}^{i} \hat{U}_j \ket{0}^{\otimes N}$.
Rather than compiling the final target state in one step, we iteratively design gate sequences preparing intermediate target states.
We maximize the fidelity between the $i$-th intermediate target state and a chosen ansatz constructed from a set of two-qubit unitaries $\{\hat{W}^i\}$ in a manner similar to what is presented in the previous section.
This gives an approximation to the $i$-th intermediate target as $\ket{\psi_{i}^\mathrm{approx}} = \prod_{j} \hat{W}^i_j \ket{\psi_{i-1}^\mathrm{approx}}$, where $\ket{\psi_{i-1}^\mathrm{approx}}$ corresponds to the compiled approximation of the previous target (or the zero state in the first step).
Here, the ansatz of two-qubit gates, $\hat{W}^i_j$, only acts on the small subset of qubits of the target unitary $\hat{U}_i$, thus reducing each variational compilation step to a small subspace of qubits.
As the rotation between intermediate states of an exact decomposition only acts on few numbers of qubits by construction, we can efficiently perform each variational compilation by tracing out all qubits the rotation does not act on.
This allows us to avoid the computation of a reverse application of the circuit to the final target state as a tensor network, thus not causing the explosion of the bond dimension observed in the variational approach.

In contrast to the QSD, the compilation of intermediate states with a classical variational approach allows for the use of different ans\"atze, which may be chosen to satisfy certain requirements about the connection topology or gate sets that can be realized on a device.
Here, we assume an all-to-all connectivity within each local block, and iteratively grow the ansatz for a unitary block until a sufficient fidelity between the target intermediate state and the ansatz is achieved.
Starting from an empty set of gates $\{\hat{W}^i\}$ initially, we add a new two-qubit gate $\hat{\Omega}$ in each step to improve the expressivity of the ansatz.
This is then optimized according to the scheme outlined in the previous section.
We repeat these two steps until the fidelity between the approximation and the intermediate target, $F[\{\hat{W}^i\}] = |\braket{\psi_{i}^\mathrm{approx}}{\psi_{i}^\mathrm{t}}|^2$, surpasses a pre-defined fidelity threshold, $\mathrm{F}^i_\mathrm{thresh}$.
To pick a new two-qubit gate to be added to the ansatz, we identify a suitable two-qubit gate by looping over all potential pairs of qubit indices and testing the fidelity improvement when adding the gate at any position in the sequence of gates.
We define the most suitable gate to be the one specified by the two qubit indices and position in the ansatz where the largest fidelity improvement is observed when initializing the unitary with the locally optimal parameters according to Eq.~\eqref{eq:local_optimal_parameterss}.

Within the protocol, we control the accuracy vs gate count trade-off in the state preparation by setting a global target fidelity for the state preparation, $F_\mathrm{target}$.
As the preparation errors in the iterative state compilation accumulate, the intermediate states generally need to be prepared to higher fidelities.
We achieve an (approximately) homogeneous distribution of the error across the ansatz by setting the threshold for the compilation of the $i$-th intermediate state to $\mathrm{F}^i_\mathrm{thresh} = \mathrm{F}^i_\mathrm{max} \times F_\mathrm{target}^{(1/N_\mathrm{blocks})}$.
Here, $N_\mathrm{blocks}$ is the total number of intermediate states which are compiled, and $\mathrm{F}^i_\mathrm{max}$ is the maximum fidelity that can be achieved for the $i$-th intermediate state, taking into account errors made in the preparation of preceding intermediate states.
The size of the blocks in the compilation is controlled by the bond dimensions in the target MPS.
Depending on the target fidelity $\mathrm{F}_\mathrm{target}$, we apply an additional bond dimension truncation of the target state to reduce the computational overhead of the method.
In analogy to the exact preparation approach, we truncate the bond dimensions of the target state $\ket{\psi_{\text{MPS}}}$ so that the fidelity between the truncated and untruncated MPS remains greater than $\sqrt{\mathrm{F}_\mathrm{target}}$.
This reduces the gate count of the compiled MPS, while ensuring that the main source of error in the state preparation circuit stems from the error made in the variational compilation of intermediate states.

\subsection{Quantum Subspace Expansion for Green's functions}
The quantum circuit optimized on the classical computer can now be executed on a quantum computer, so that $\ket{\psi_{\text{QC}}}$ is produced on the quantum computer. We use it as starting point to perform its time evolution on the quantum computer and with it obtain the GF. There are several quantum algorithms that use the time evolution of quantum states to compute the GF on a quantum computer~\cite{troyer,PhysRevX.8.041015,Berry2018,Jaderberg2020,PhysRevResearch.2.033281,Keen2020,2sites,mlde,Keen2020,PhysRevB.105.115108,kvqa,qseg}, which differ in scalability and required quantum resources.  Here, we use a modified version of the quantum subspace expansion algorithm for GFs (QSEG)~\cite{qseg} due to its potential scalability to large systems with a moderate number of required Trotter steps. Compared to the QSEG algorithm in Ref.~\cite{qseg}, here we obtain the GS wave function directly on the quantum computer using the MPS based circuit outlined above. This significantly reduces the number of matrix elements that need to be evaluated on a quantum computer when compared to Ref.~\cite{qseg}.

Within QSEG the greater GF in Eq.~\ref{eq:greater} is represented using a Krylov basis set with a continued fraction
\begin{equation}
  G^>(z) = \frac{1}{z - a_0-\frac{b_1^2}{z - a_1-\frac{b_2^2}{z -  a_{2}....}}},
  \label{eq:continued_fraction}
\end{equation}
and analogously for the lesser GF.
The $a_{i }$ and $b_{i}$ are obtained using the Lanczos algorithm, which orthogonalizes the Krylov basis, defined as $\ket{\phi_0},H\ket{\phi_0},H^2\ket{\phi_0}...$, where $\ket{\phi_0} = \hat{c}^\dd\ket{\mathrm{GS}}$. Since the creation operator is a sum of two Pauli strings, the state $\ket{\phi_0}$ is also a sum of two wave vectors, one for each of these Pauli strings~\cite{qseg}.
Once orthogonalized within the Lanczos scheme, the Hamiltonian is tridiagonal in this basis, bringing the GF in Eq.~\ref{eq:greater} into the continued fraction representation presented in the equation above.

To represent the Krylov basis, we use the QSEG framework of Ref.~\cite{qseg}, where each Krylov vector is decomposed as a linear combination of subspace basis vectors $\ket{\psi_{k}} = \hat{U}(\Delta_t)^k \ket{\phi_{0}}$, where $\hat{U}$ is an unitary operator approximating the time evolution $e^{-i\Delta_t H}$ with a Trotter expansion (for ease of notation we absorb $\hbar$ into our definition of time).
Here  $k \in [-n_l,n_l]$, where $n_l$ is an integer that sets the size of the basis used to expand the Krylov vectors~\cite{qseg}.
One then needs to evaluate on the quantum computer the overlap and Hamiltonian matrix elements for these basis vectors, given by
\begin{align}
  S_{ij} & = \bra{\phi_{0}}\hat{U}(\Delta_t)^{-i}\hat{U}(\Delta_t)^j\ket{\phi_{0}},\label{eq:S_matrix}        \\
  H_{ij} & = \bra{\phi_{0}} \hat{U}(\Delta_t)^{-i}\hat{H}\hat{U}(\Delta_t)^j\ket{\phi_0}, \label{eq:H_matrix}
\end{align}
respectively.
These can be obtained on a quantum computer with different methods, such as for example the Hadamard test~\cite{filip2022variational}.
To perform the Trotter expansion, we use the scheme in Ref.~\cite{qseg}, where the time evolution of the quadratic part of the Hamiltonian is performed exactly.
The evaluation of these matrix elements on the quantum computer is schematically illustrated in the third box in Fig.~\ref{fig:schematic}.
We note that one may greatly reduce the number of matrix elements that needs to be computed by exploiting the fact that $S_{ij}$ is a Toeplitz matrix, where the elements only depend on $i-j$, and that, if a small enough Trotter step is used, then also $H_{ij}$ becomes approximately a Toeplitz matrix (see Appendix~\ref{appendix:commutation} for details).

These $S$ and $H$ matrices obtained on a quantum computer are then used on a classical computer to perform the orthogonalized Krylov basis construction~\cite{qseg}, and with it to obtain the GF and the DOS.
This is schematically illustrated in the last panel in Fig.~\ref{fig:schematic}.

\section{Benchmarking the pipeline for a strongly-correlated material}
\begin{figure}
  \centering
  \includegraphics[width=\columnwidth]{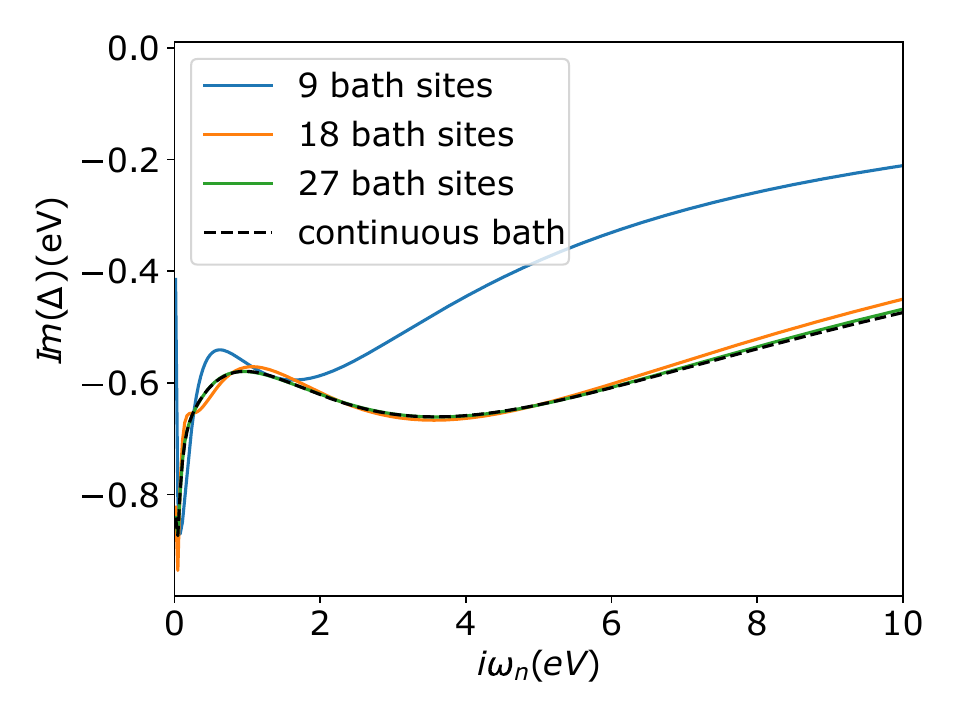}
  \caption{Comparison between the target continuous bath hybridization function (black dashed line) and the hybridization function for a discretized bath with increasing number of bath sites, on the imaginary axis ($\omega_n=(2 n + 1)\pi/\beta_\mathrm{f}$).}
  \label{fig:fit}
\end{figure}

\begin{figure*}
  \centering
  \includegraphics[]{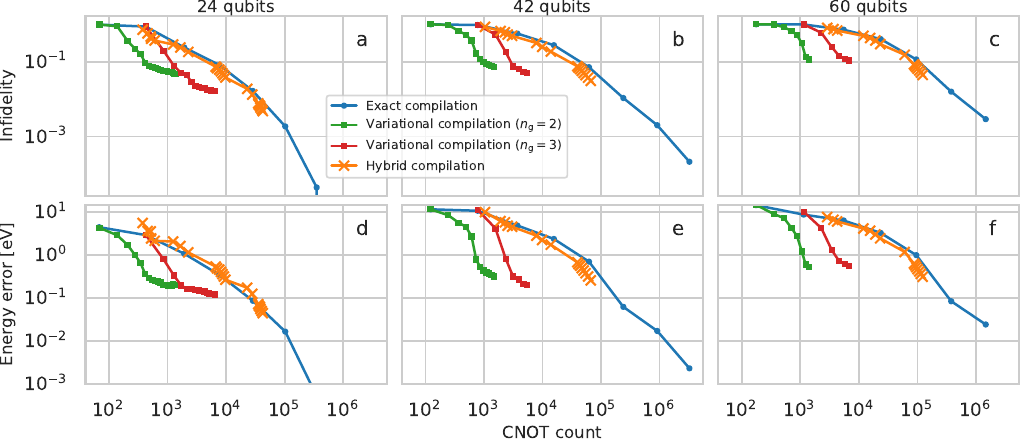}
  \caption{Infidelity (upper panels), and energy difference (lower panels) of the compiled state preparation quantum circuits with respect to the number of CNOT gates for the \srvo~GS
    with 3 impurity sites, for 9 (left), 18 (center), and for 27 bath orbitals (right). The figure shows the results for a compilation of a target ground state MPS with an exact compilation of a truncated MPS (blue, circles), a variational compilation into staircase layers with $N_g=2$ (green squares) and $N_g=3$ (red squares) qubit gates, and a hybrid approach (orange crosses).}
  \label{fig:srvo3_nlayer}
\end{figure*}

We demonstrate our algorithm on the prototypical strongly correlated metal with perosvskite structure \srvo\ \cite{PhysRevB.73.155112,PhysRevLett.115.256402}, where a non-perturbative method such as DMFT is required to obtain the correct renormalization of the quasi-particle bands.
We first compute the ground state using density functional theory within the local density approximation with the Questaal code~\cite{questaal}.
Then, the DMFT subspace is constructed via local projection onto the three V t$_{2g}$ orbitals.
The continuous bath is discretized by fitting the hybridization function, $\Delta$, with a finite number of bath orbitals on the Matsubara axis for a fictitious inverse temperature of $\beta_{\text{f}} {=} 200 $ eV$^{-1}$.
The finite $\beta_{\text{f}}$ broadens sharp peaks in the hybridization, so that it can be computed with a finite grid on the imaginary axis.
In Fig.~\ref{fig:fit}, we plot the obtained hybridization function of the first DMFT iteration with 9, 18 and 27 bath orbitals on the Matsubara axis, and compare it to the target continuous bath hybridization.
There is a significant discrepancy between the nine bath orbitals results and the continuous bath solution.
The agreement improves for 18 bath orbitals, and the representation becomes effectively exact for 27 bath sites.

We therefore perform the computation of the GS for the three impurity sites coupled to 9, 18, and 27 bath sites, requiring 24, 42, and 60 qubits in total, respectively, when using the Jordan-Wigner transform and two spins per site.
We make use of a set of orbitals for the bath that make the hopping matrix (containing the hopping matrices for the impurity, $\e_{ij}$, and bath, $\e^d_{ij}$, orbitals and the hybridization matrix, $V_{ij}$) block tridiagonal, where each block has a dimension of $n_\mathrm{imp}$.
We perform a DMRG calculation to obtain an accurate MPS representation of the GS wave function, $\ket{\psi_\mathrm{MPS}}$, and GS energy, $E_\mathrm{GS}$.
This results in MPS GS representations with maximum bond dimensions of $\chi_\mathrm{max} = 245$ (24 qubits), $\chi_\mathrm{max} = 745$ (42 qubits), and $\chi_\mathrm{max} = 810$ (60 qubits).
While the ground state problem is thus still classically manageable, the order of magnitude of the bond dimensions also indicates that the ability to capture a time-evolution, typically causing an increase in the bond dimension, will mostly be classically intractable with tensor network techniques.
This is where the application of quantum computing offers an appealing alternative as all the circuits to perform a time evolution can be obtained easily.

We set up the quantum circuits to obtain $\ket{\psi_\mathrm{QC}}$ based on the introduced compilation protocols. Fig.~\ref{fig:srvo3_nlayer} presents the convergence of the quantum circuit to the MPS state for different gate counts of the quantum circuit.
The top panel presents the convergence in terms of the infidelity, $1-F$, for calculations at different system sizes, and the bottom panel shows the corresponding energy difference, $\Delta E$, of the optimized quantum circuit as a function of the number of CNOT gates in the circuits transformed into single qubit rotations and CNOT gates via the QSD.
In general, all three compilation protocols give a consistent improvement in the representation of the state for more complex quantum circuits, which manifests in smaller infidelities and energy errors as the CNOT counts increase.

All three compilation approaches are viable alternatives to prepare an approximate ground state on the device.
The practical suitability of the techniques depends on the specifics of the performed experiment, in particular the accuracy in the GS preparation necessary for the computation of Green's functions, as well as hardware requirements of the quantum computing device.
While the exact compilation approach guarantees that the MPS can be compiled to arbitrary accuracies, this comes at the cost of a gate count which is linear in system size and in bond dimension of the MPS.
This generally results in large CNOT counts in the extracted quantum circuit.
Already for the 24 qubit system, more than $10^6$ CNOT gates are required for the exact state preparation, making the practical execution of the approach on near-term hardware infeasible.

Here is where the variational compilation offers an appealing alternative.
With a variational compilation using staircase layers with $n_g=2$, we obtain infidelities in the prepared state of the order of $1-F = 10^{-1}$ using significantly fewer CNOT gates than in the exact compilation, requiring only $\approx 400$ CNOT gates for the compilation of the 24 qubit GS to this level of accuracy.
However, despite a rapid initial convergence, we are not able to reach an energy difference below $10 \, \mathrm{meV}$ with the variational circuit optimization by increasing the gate count of the variational ansatz.
In the 42 qubit case, we can obtain an overlap of $\approx 92\%$ with 2-qubit gates (which corresponds to an error in the energy of $\approx 0.3 \, \mathrm{eV}$) before it plateaus and the classical optimization of the ansatz becomes infeasible.
While we are able to reduce this to an energy error of $\approx 0.2 \, \mathrm{eV}$ by utilizing a staircase ansatz with $n_g=3$, at the cost of increasing the CNOT count to $\approx 5000$, this also runs into computational limitations if a higher accuracy state preparation is desired.

If the number of CNOT gates is not a major limitation of the device, which may occur in the fault-tolerant era, building upon the exact decomposition of the MPS as a quantum state offers an alternative to achieve higher state preparation accuracies. In the studied example, we obtain a comparable relationship between the CNOT count of the circuit and the preparation accuracy for a variational compilation of intermediate target states and their exact preparation via the QSD.
Although the hybrid approach does not significantly reduce the number of CNOT gates in the preparation circuit, we expect its main advantage to lie in the fact that the variational circuits for the representation of intermediate states can easily be informed by hardware requirements, as is usual for variational approaches.
This may, for example, help to reduce the depth of the circuits by going to hardware-efficient ansatzes~\cite{Leone_2024} for the different blocks, or base the ansatzes on connectivity topologies of the device.
Already within the employed protocol, not constraining the connectivity in the ansatz or targeting low-depth circuits, we observe a small but consistent reduction in the depth of the circuits as compared to the QSD.
This is highlighted in the appendix, where we plot the same data as in Fig.~\ref{fig:srvo3_nlayer}, but as a function of the circuit depths (measured as the number of non-parallel CNOT gates).
In order to reach a fidelity of $\approx 0.9$ for the preparation of the 60 qubit state, the exact compilation leads to a depth of $\approx 10^5$ gates, which is reduced by $\approx 20 000$ gates with the hybrid approach.

While the exact compilation approach currently offers the only route to reliably achieve arbitrary state preparation accuracies, in the following, we provide numerical evidence that the algorithm can even succeed with a non-exact compilation of the target state, as achieved by the variational approach.
The ability to use the sparse and compact state preparation schemes can thus significantly reduce the resource requirements for a realization on actual hardware. 

\begin{figure*}
  \centering
  \includegraphics[]{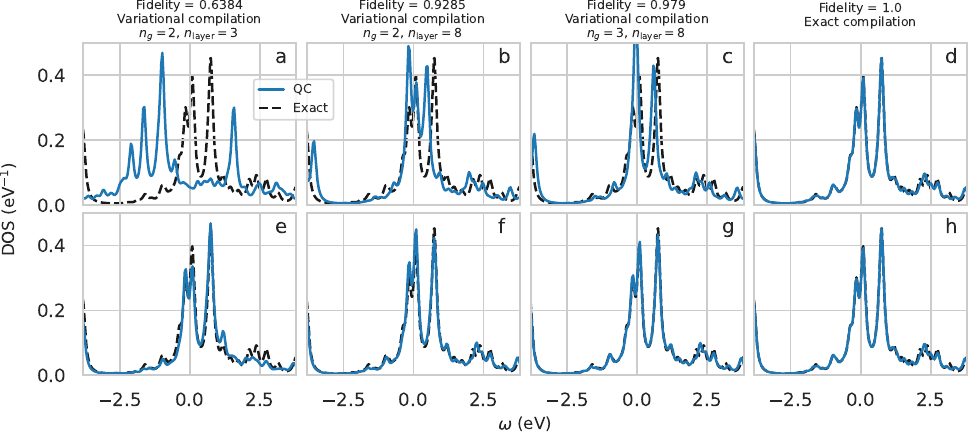}
  \caption{
    The local DOS of the first DMFT iteration for \srvo~computed with the MPS+QSEG algorithm for $n_l = 200$ and $\D_t=0.05$ eV$^{-1}$. Here we show the DOS of the $t_{\text{2g}}$ $xy$ orbital (the two other $t_{\text{2g}}$ orbitals have very similar accuracy). The upper panels (a-d) show the GF computed with the approximate GS energy obtained on the quantum circuit ($E_{\text{QS}} = \expval{\hat{H}}{\psi_{\text{QC}}}$), and the lower panels (e-h) show the results for the accurate MPS GS energy ($E_{\text{GS}} = \expval{\hat{H}}{\psi_{\text{MPS}}}$). In each panel we compare the DOS obtained using the quantum circuit representation of the GS (blue solid lines) with the DOS obtained using an exact diagonalization technique (black dashed lines). The GS overlap $f$ for each DOS is given above the panels, as well as the used values of $n_\mathrm{g}$ and $n_\mathrm{layer}$ to obtain this overlap.}
  \label{fig:fidelity_gf}
\end{figure*}

Having obtained the quantum circuit for the GS wave function using the TN classical computing approach above, we can execute the circuit on a quantum computing emulator. With this $\ket{GS}=\ket{\psi_\mathrm{QC}}$ we then evaluate the matrix elements in Eqs.~\ref{eq:S_matrix} and~\ref{eq:H_matrix} by performing the Trotter time evolution, and with them obtain the GF (see schematic~\ref{fig:schematic}). We use the Qulacs state vector emulator~\cite{qulacs} to perform these simulations. Since the computing resources required in these state vector emulations are much higher than for our MPS based GS computations, we can only simulate the 24 qubit system, while the 42 and 60 qubit systems are beyond what can be treated with our available resources.
We use a Trotter step of $\Delta_t = 0.05 \, \mathrm{eV}^{-1}$ and $n_l = 200$ for the basis used in the QSEG algorithm.

We systematically evaluate the quality of the obtained GF for different levels of fidelity $F=|\braket{\psi_{\text{MPS}}}{\psi_{\text{QC}}}|^2$. To this aim, we compute the exact reference GF using an exact diagonalization technique~\cite{edcode,mlde}. In Figs.~\ref{fig:fidelity_gf}a-d, we compare the local DOS ($\mathrm{DOS}(\omega)= -\frac{1}{\pi} \text{Im}G(\w + i\d)$, where here we use $\delta=0.1$ eV) obtained using the compiled quantum circuit representations of the MPS GS with increasing overlap. The exact compilation of the GS MPS sets the baseline accuracy for the DOS computed with the subspace approach with a state preparation fidelity of $F=1$ shown in Fig.~\ref{fig:fidelity_gf}d. We find that, only for this exact compilation of the GS, the DOS is in approximate agreement with the exact reference DOS. While this shows that the algorithm is able to faithfully extract the DOS with the chosen parameters, it may also suggest that very high fidelities, and hence accurate GS energies, are required for the quantum circuit representation of the GS wave function.

To obtain an accurate GF also for lower values of the overlap, we note that a significant part of the difference from the exact DOS is caused by shifts in the peaks. One reason for such shifts is that the GS energy computed with the quantum circuit, $E_{\text{QC}}=\expval{\hat{H}}{\psi_{\text{QC}}}$, differs from the exact $E_\mathrm{GS}$, as shown in Fig.~\ref{fig:srvo3_nlayer}. To improve the GF obtained with our quantum algorithm we therefore take advantage of the fact that we have access to the exact MPS GS energy $E_{\text{GS}}$ independently of its quantum circuit representation, and use this energy $E_\mathrm{GS}$ in Eqs.~\ref{eq:greater} and~\ref{eq:lesser} for the computation of the GFs. The results when using this approach are shown in Figs.~\ref{fig:fidelity_gf}e-h. One can see that with this method the DOS converges to the exact DOS already using the variationally compiled state with $n_\mathrm{layer}=8$ layers of $n_g =2$ qubit gates, giving a lower fidelity of about $F=0.93$.
This is an important advantage of our scheme: we do not need to have a perfect match between the exact GS and the compiled quantum circuit, because the exact GS energy is obtained with excellent numerical precision on a classical computer. For the 42 and 60 qubit calculations, we obtain fidelities of up-to $\approx 93 \%$ and $\approx89\%$ using the resource-efficient variational circuits constructed from two-qubit staircase layers (see Fig.~\ref{fig:srvo3_nlayer}). We expect this level to be sufficient to extract an accurate representation of the GF if the time-evolution could be executed, either on an emulator with larger classical computing resources, or directly on a quantum computer.

The concept presented here of preparing the initial state on the quantum computer from a tensor network solution obtained on a classical computer, followed by a Hamiltonian time evolution on the quantum computer, also applies to similar tasks. For example, when the tensor network approaches themselves give only an approximate ground state, one can first use the tensor network based approach to obtain a variational approximation of the ground state that is initially prepared on the quantum computer, and then perform quantum phase estimation in analogy to a time-evolution to improve it and bring it closer to the true ground state~\cite{Sayginel_TFVQE_2023,berry2024rapidinitialstatepreparation}.

\section{Conclusion}
In conclusion, in this article we have proposed an Anderson impurity solver using classical and quantum computing resources. The GS is first determined with classical TN methods, and then a quantum circuit representing this GS is constructed on the classical computer. This circuit can be executed on a quantum computer, and subsequently time evolution of the state can be performed to obtain the Hamiltonian and overlap matrix elements within a quantum subspace expansion approach, and with these the GF is obtained. Combining the strengths of the classical and quantum methods allows us to obtain the GF of an impurity model for the real material \srvo, represented on 24 qubits on an emulator. We outline different strategies to prepare the initial state for the same material with 42 and 60 qubits that could allow for systematic improvement of the constructed GF if either larger classical computing resources would be available for the time evolution quantum emulation, or if the time evolution was executed directly on a quantum computer.
To further improve the scalability of the classical tensor network computation, this method can be extended to more general tree tensor networks or fork tensor networks, which have shown success for treating multi-orbital impurity models~\cite{ttn1, FTN1}.

\subsection*{ Acknowledgments}
The authors acknowledge the support of the UK government department for Business, Energy and Industrial Strategy through the UK National Quantum Technologies Programme. FJ acknowledges support through an Explorers Award from the National Physical Laboratory’s Directors’ Science and Engineering Fund. EF acknowledges the support of an industrial CASE (iCASE) studentship, funded by the UK Engineering and Physical Sciences Research Council (grant EP/T517665/1), in collaboration with the university of Strathclyde, the National Physical Laboratory, and Quantinuum. L. P. L, Y.R. and I. R. acknowledge support from
the Engineering and Physical Sciences Research Council [grant EP/Y005090/1].

\bibliographystyle{apsrev4-2}
\bibliography{bibli}

\appendix

\section{CNOT count of $n_\mathrm{g}$-qubit Gates \label{appendix:CNOT}}

\begin{figure*}[htb!]
  \centering
  \includegraphics{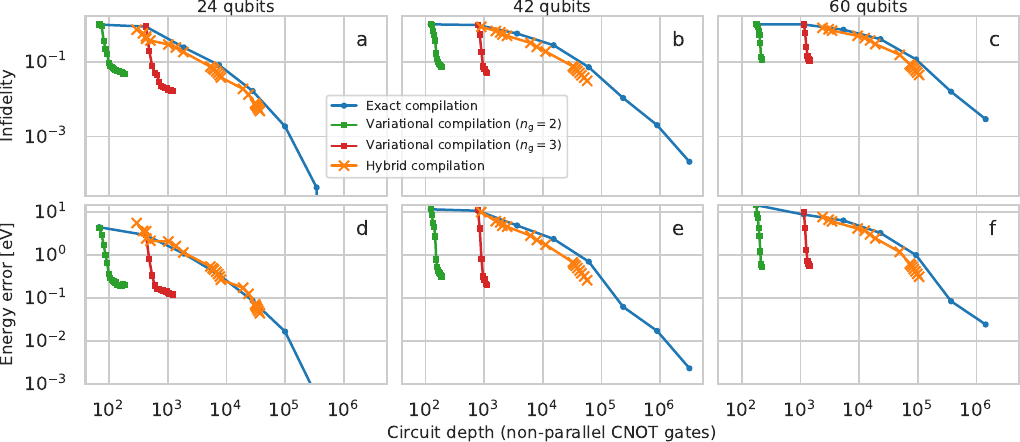}
  \caption{Same data as plotted in Fig.~\ref{fig:srvo3_nlayer}, where the horizontal axis now is the circuit depth measured as the number of non-parallel CNOT gates in the circuit.}
  \label{fig:srvo3_cnot_depth}
\end{figure*}

An arbitrary $n_\mathrm{g}$-qubit unitary operator can be implemented on a quantum computer using one-qubit gates ($R_x$, $R_y$, $R_z$) and CNOT gates. Table~\ref{tab:CNOTcount} provides the number of CNOT gates required to implement one general $n_\mathrm{g}$-qubit gate using the optimized quantum Shannon decomposition, which is given by $N_\text{CNOT}=\frac{23}{48}4^{n_g}-\frac{3}{2}2^{n_g}+\frac{4}{3}$~\cite{Shende_2006}, for $n_\mathrm{g}$ up to 5.
\begin{table}[H]
  \centering
  \begin{tabular}{|c|c|}
    \hline
    Number of qubits & CNOT gate count \\
    \hline
    2                & 3               \\ \hline
    3                & 20              \\ \hline
    4                & 100             \\\hline
    5                & 444             \\ \hline
  \end{tabular}
  \caption{CNOT count of a $n_\mathrm{g}$-qubit gate using the optimized quantum Shannon decomposition.~\cite{Shende_2006}}
  \label{tab:CNOTcount}
\end{table}

Some of the unitaries in the staircase ansatz can be performed in parallel which reduces the depth - in terms of non-parallel CNOT gates - of the circuit to $((n_{\mathrm{l}} - 1)n_{\mathrm{g}} + (n_{\mathrm{q}} - n_{\mathrm{g}} + 1))N_{\textrm{CNOT}}(n_{\mathrm{g}})$, with $n_{\mathrm{l}}$ the number of layers and $n_{\mathrm{q}}$ is the total number of qubits.
Fig.~\ref{fig:srvo3_cnot_depth} presents the same data as shown in Fig.~\ref{fig:srvo3_nlayer}, namely the infidelity for the 24, 42, and 60 qubit calculations in the top panel, and the corresponding energy difference in the bottom panel, but now as a function of the number of non-parallel CNOT gates.
In terms of the depth, the variationally optimized circuit of 2 qubit unitaries is generally far more efficient than the other schemes, achieving fidelity levels of the exact decomposition with almost a factor of $10^3$ less gates.
As outlined in the main text, the practically achievable preparation fidelity is however limited, requiring the application of alternative compilation strategies if the initial state needs to be prepared to higher fidelities.

\section{Reducing the number of matrix elements that needs to be computed \label{appendix:commutation}}

For each matrix $S_{ij}$ and $H_{ij}$ (Eqs.~\ref{eq:S_matrix} and~\ref{eq:H_matrix} in the main text) there are  $(2n_l+1)^2$ matrix elements. Both matrices are symmetric, reducing the number of independent elements to $\frac{(2n_l+1)(2n_l+2)}{2}$. Moreover, $S_{ij}$ is a Toeplitz matrix (its elements only depend on $i-j$), further significantly reducing the number of its independent elements to only $2n_l+1$. If the time evolution operator $\hat{U}(\Delta_t)$ is executed exactly, one has $H_{ij} = \bra{\phi_0} \hat{H}\hat{U}(\Delta_t)^{j-i}\ket{\phi_0}$, since in this case $\hat{H}$ commutes with $\hat{U}(\Delta_t)=e^{-i\Delta_t H}$, making $H_{ij}$ a Toeplitz matrix as well.

In practice, however, $\hat{U}(\Delta_t)$ is executed approximately, hence $\hat{H}$ only approximately commutes with $\hat{U}(\Delta_t)$.
More precisely, in this work the time evolution operator is approximated by the Trotter expansion $\hat{U}(\Delta_t) \approx e^{-i\hat{H}_\text{int}\Delta_t/2}e^{-i\hat{H}_{0}\Delta_t}e^{-i\hat{H}_\text{int}\Delta_t/2}$, where the error scales as $\mathcal{O}(\Delta_t^3)$. Here we have introduced
\begin{align}
  H_0 =            & \sum^{n_{\text{imp}}}_{ij}\sum_{\s} \e_{ij} \hat{c}^\dd_{i\s} \hat{c}_{j\s} +  \sum_\s \sum^{n_b}_{ij}\e^d_{ij} \hat{d}^\dd_{i\s}\hat{d}_{j\s}\nonumber \\ +& \sum^{n_{\text{imp}}}_{i}\sum^{n_b}_{j}\sum_\s V_{ij} (\hat{c}^\dd_{i\s} \hat{d}_{j\s}+ \hat{d}_{j\s}^\dd \hat{c}_{i\s})\, ,\\
  H_{\text{int}} = & \sum^{n_{\text{imp}}}_{ijkl }  \sum_{\s \s'} U_{ijlk} \hat{c}^\dd_{i\s} \hat{c}^\dd_{j\s'} \hat{c}_{k\s'} \hat{c}_{l\s}.
\end{align}

Therefore, as $\Delta_t$ decreases, the error introduced by performing the commutation $H_{ij} = \bra{\phi_0} \hat{U}(\Delta_t)^{-i} \hat{H}\hat{U}(\Delta_t)^{j}\ket{\phi_0} \approx \bra{\phi_0} \hat{H}\hat{U}(\Delta_t)^{j-i}\ket{\phi_0}$ also decreases, until $H_{ij}$ becomes a Toeplitz matrix to a good approximation. However, decreasing $\Delta_t$ also increases the depth of the circuit required to reach a given total evolution time. Therefore it needs to be verified whether performing the commutation leads to a reasonable error for practical values of $\Delta_t$.

\begin{figure}
  \centering\vspace{1em}
  \includegraphics[]{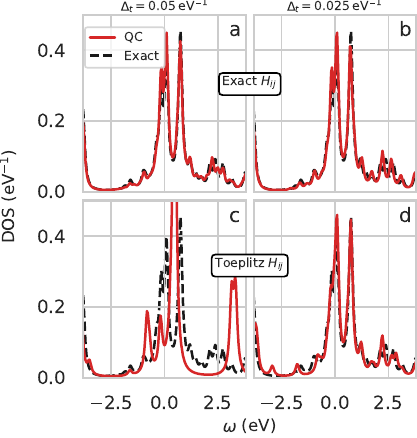}
  \caption{The DOS of the impurity model calculated using MPS+QSEG (orange curves) for $\Delta_t=0.05$ eV$^{-1}$ (left panels) and $\Delta_t=0.025$ eV$^{-1}$ (right panels) using $n_T=4$ Trotter steps per Krylov basis state, compared with the exact diagonalization result (dashed black line). The upper panels show the DOS obtained using $H_{ij}$ as defined in Eq.~\ref{eq:H_matrix}, while the DOS shown in the bottom panels is obtained using the approximation $H_{ij} \approx \bra{\phi_0} \hat{H}\hat{U}(\Delta_t)^{j-i}\ket{\phi_0}$.}
  \label{fig:GF_appendix}
\end{figure}
We test the resulting accuracy of the GF for our \srvo\ AIM system when using a Toeplitz approximation for $H_{ij}$.
To ensure systematic reduction of the Trotter error for a fixed Krylov basis set size, we employ a multi-step approach with multiple Trotter steps per Krylov basis state.
This gives $H_{ij} \approx \bra{\phi_0} \hat{H}\hat{U}(\Delta_t/n_T)^{n_T(j-i)}\ket{\phi_0}$, where $n_T$ is the number of Trotter steps applied per basis state.
We test this approximation for two different values of $\Delta_t$, namely $\Delta_t=0.05$ eV$^{-1}$ and $\Delta_t=0.025$ eV$^{-1}$, using a fixed $n_T = 4$.
We keep the number of basis states in the subspace expansion equal to $n_l=200$ for both cases, such that the depth of both circuits is the same. The other parameters are the same as those used for the DOS plot in Fig.~\ref{fig:fidelity_gf}h.

The upper panels of Fig.~\ref{fig:GF_appendix} show the DOS obtained using $H_{ij}$ as defined in Eq.~\ref{eq:H_matrix}, while the DOS shown in the bottom panels is obtained using the approximation $H_{ij} \approx \bra{\phi_0} \hat{H}\hat{U}(\Delta_t)^{j-i}\ket{\phi_0}$.
For $\Delta_t = 0.05$  eV$^{-1}$, which is the value also used in the calculation in the main text for Fig.~\ref{fig:fidelity_gf}h (however with $n_T =1$), the use of the approximated $H_{ij}$ gives a DOS that is not in agreement with the exact diagonalization result (dashed black line).
This points to the accumulation of significant Trotter errors in the circuit for this Trotterization of the time evolution, which does not affect the DOS when evaluating the matrix $H_{ij}$ exactly (see Fig.~\ref{fig:GF_appendix}a).
However, when decreasing the timestep to $\Delta_t = 0.025$  eV$^{-1}$, the agreement between the obtained DOS with the exact solution is significantly better, both with and without the Toeplitz approximation to the $H_{ij}$ matrix.
These results confirm that the approximated relation $H_{ij} \approx \bra{\phi_0} \hat{H}\hat{U}(\Delta_t)^{j-i}\ket{\phi_0}$ can be applicable in practical calculations provided that $\Delta_t/n_T$ is small enough, in which case the number of matrix elements to be computed is significantly reduced by taking advantage of the Toeplitz form of the matrix.
Whether for a general system it is feasible to reach the small $\Delta_t/n_T$ required depends on the details of the system, the characteristics of the quantum device, and the construction of the time evolution circuit, and must be assessed on a case-by-case basis.

\end{document}


\title{Supplementary material: Quantum subspace expansion algorithm for Green's functions}
\begin{abstract}

\end{abstract}

\author{Fran\c{c}ois Jamet}
\email{francois.jamet@npl.co.uk}
\author{Abhishek Agarwal}
\author{Ivan Rungger}
\email{ivan.rungger@npl.co.uk}
\affiliation{National  Physical  Laboratory,  Teddington,  TW11  0LW,  United  Kingdom}

\maketitle
\section{Computation of matrix elements of $H$ and $S$}
In this section, we present how to compute matrix elements  $S_{ij}=\bra{\phi_0}\hat{U}^\dd_i\hat{U}_j\ket{\phi_0}$ and $H_{ij}=\bra{\phi_0}\hat{U}^{\dd}_i\hat{H}\hat{U}_j\ket{\phi_0}$ for the ground state (Eqs. (2) and (3) in the main text), and  $(\bm{S}_{\psi})_{ij}=\bra{\phi_0}\hat{U}^{\dd}_i \hat{\s}_a \hat{U}^{\dd}_j\hat{U}_k \hat{\s}_b\hat{U}_l\ket{\phi_0}$ and $(\bm{H}_{\psi})_{ij}=\bra{\phi_0}\hat{U}^{\dd}_i \hat{\s}_a \hat{U}^{\dd}_j\hat{H}\hat{U}_k \hat{\s}_b\hat{U}_l\ket{\phi_0}$ for the Krylov states (Eqs. (16) and (20) in the main text). Here $\hat{\s}_a$ stands for a Pauli string of the form $\prod_{k=0}^{a-1}\s^{k}_{Z}\s^{a}_{X/Y}$ where $\s^k_{X/Y/Z}$ is a Pauli X, Y or Z operator of qubit k. In general, these matrices may be computed using a Hadamard test \cite{Stair2020,1909.08925}, which requires deep circuits including Toffoli gates.

An alternative approach, which avoids the use of Hadamard tests, has been proposed in Refs. \cite{Kyriienko2020,cortes2021quantum}. The matrix elements of the form $\bra{\phi}\h U_i^{\dd} \h U_j\ket{\phi}$, where $\hat{U}_i, \hat{U}_j$ are unitary operators and $\ket{\phi}$ is a quantum state, can be computed without a Hadamard test by using the so-called multi-fidelity estimation protocol, if one can find a reference state $\ket{R}$ such that:
\begin{itemize}

\item Condition I: $\bra{R}\h U_i^{\dd} \h U_j\ket{R}=r_R e^{i\theta_R}$ is known
\item Condition II: $\bra{R}\h U_i^{\dd} \h U_j\ket{\phi}=0$
\item Condition III: $\ket{\phi}+\ket{R}$ can be constructed.
\end{itemize}
Here $r_R$ is a real number and $\theta_R$ describes the complex phase. Under these conditions we can compute the following state fidelities on a quantum computer:
\begin{align}
  F_1 =&   |(\frac{\bra{\phi} + \bra{R}}{\sqrt{2}})\h U_i^{\dd} \h U_j (\frac{\ket{\phi} + \ket{R}}{\sqrt{2}}) |^2  \nonumber \\=&  \frac{1}{4}|\bra{\phi}\h U_i^{\dd} \h U_j \ket{\phi} +  \bra{R}\h U_i^\dd \h U_j  \ket{R} |^2,\\
F_2=&|\bra{\phi}\h U_i^{\dd} \h U_j \ket{\phi}|^2.
\end{align}
An overview of different methods to compute such state fidelities is presented in Ref. \cite{cortes2021quantum}.
With these equations, one can then obtain $\bra{\phi}\h U_i^{\dd} \h U_j \ket{\phi}=r e^{i\theta}$ as
\begin{align}
  r & = \sqrt{F_2},\\
  \theta & =\arccos{}\Bigg(\frac{4F_1-F_2-r^2_R}{2r_R \sqrt{F_2} }\Bigg) + \theta_R.
\end{align}

We will now outline how the formalism outlined above can be used to compute the matrix elements of interest for the QSEG when the total number of particles is a symmetry of the Hamiltonian.

\subsubsection{Computation of: $S_{ij}=\bra{\phi_0}\hat{U}^\dd_i\hat{U}_j\ket{\phi_0}$}
\label{sec:S}
In the main text we construct the basis for the ground state and the Krylov basis with operators $\hat{U}_i,\hat{U}_j$, which are  Suzuki-Trotter expansions of the Hamiltonian $\h H$, as outlined in Eq. (22) in the main text. If $\h H$ conserves the total number of particles $\hat N$, i.e $[\h H,\h N]=0$,  then $U_i$ and $U_j$ also conserve the number of particles. In this case, condition II can be fulfilled if  $\ket{\phi_0}$ and $\ket{R}$  are eigenvectors of $\h N$ and if they have different number of particles. Here, we choose $\ket{R}$ to be the state with no particles (or all particles in the special case where $\ket{\phi_0}$ contains no particles). With this choice of $\ket{R}$, condition I is also fulfilled,
since  $\hat{U}^\dd_i\hat{U}_j\ket{R}$ has no particles as well, so that it is equal to $\ket{R}$ up to a phase.  This phase is equal to $-h_0 T$, where $T$ is the time used in the Suzuki-Trotter expansion and $h_0$ is the constant term of the Hamiltonian written with fermionic operators. In Eq. 29 in the main text, there is no constant term so the phase is zero.

Finally, condition III depends on the construction of the reference state and is explained in Sec. \ref{sec:HF_prepa} of this document.

\subsubsection{Computation of: $H_{ij}=\bra{\phi_0}\hat{U}^\dd_i\h H \hat{U}_j\ket{\phi_0}$}
\label{sec:H}
For the construction of the matrix elements $H_{ij}$, $\h H$ is decomposed as $\h H {=} \sum_k h_k \h P_k$, where $h_k$ is a coefficient and $P_k$ a Pauli string. Some of the Pauli strings do not conserve the total number of particles, so that $[\h P_k, \h N] \neq 0$. In  the AIM, the terms such as $c_k^\dd c_l+ c_l^\dd c_k=\frac{1}{2} (\hat{\s}^X_k \prod_{k<m<l} \hat{\s}^Z_m\hat{\s}^X_l+\hat{\s}^Y_k\prod_{k<m<l} \hat{\s}^Z_m\hat{\s}^Y_l)$  generate such Pauli strings (for ease of notation, in the rest of the section we drop the Pauli Z strings, which do not change the number of particles). If $\ket{a}$ and $\ket{b}$ are eigenstates of $\h N$, and $|\expval{\h N}{a}-\expval{\h N}{b}|>2$, then $\bra{a}\hat{\s}^{X/Y}_i\hat{\s}^{X/Y}_j\ket{b}=0$, since $\hat{\s}^{X/Y}_i$ connects only adjacent total number of particles sectors.
Therefore, if $\expval{\h N}{\phi_0}>2$, we can choose $\ket{R}=\ket{0} \otimes \ket{0}...\otimes\ket{0}$, while in the special case where if $\expval{\h N}{\phi_0}\leq 2$, we can choose $\ket{R}=\ket{1} \otimes \ket{1}...\otimes\ket{1}$. The motivation behind these choices is to have an $\ket{R}$ state that can be easily obtained in a quantum circuit, and where condition I can be fulfilled.
Finally, condition III depends on the construction of the reference state, which is explained in Sec. \ref{sec:HF_prepa} of this document.

\subsubsection{Computation of: $({S}_{\psi})_{ij}=\bra{\phi_0}\hat{U}^{\dd}_i \hat{\s}_a \hat{U}^{\dd}_j\hat{U}_k \hat{\s}_b\hat{U}_l\ket{\phi_0}$}

For $(\bm{S}_{\psi})_{ij}$, the choice of $\ket{R}$ is the same as in Sec. \ref{sec:H}, since there are 2 Pauli operators which change the total number of particles.

However, more care needs to be taken in condition II, which involves the computation of $\bra{R}\hat{U}^{\dd}_i \hat{\s}_a \hat{U}^{\dd}_j\hat{U}_k \hat{\s}_b\hat{U}_l\ket{R}$. $\hat{U}_l\ket{R}$ is equal to $\ket{R}$ up to a phase $\theta_l$, such that $\bra{R}\hat{U}^{\dd}_i \hat{\s}_a \hat{U}^{\dd}_jU_k \hat{\s}_bU_l\ket{R}{=} e^{i(\theta_l-\theta_i)}\bra{R} \hat{\s}_a \hat{U}^{\dd}_jU_k \hat{\s}_b\ket{R}$. To compute $\bra{\phi_1}\hat{U}^{\dd}_jU_k\ket{\phi_2} $ with $\ket{\phi_1}=\hat{\s}_a\ket{R}$ and $\ket{\phi_2}=\hat{\s}_b\ket{R}$, we can generalize the method presented in Sec. \ref{sec:S} with $\ket{\phi_1}\neq \ket{\phi_2}$. This can be done using the same technique as long as we have $\bra{\phi_1}\ket{R} {=}\bra{\phi_2}\ket{R}{=}0$.

In this case, we have
\begin{align}
  F_1 =&   |(\frac{\bra{\phi_1} + \bra{R}}{\sqrt{2}})\h U_i^{\dd} \h U_j (\frac{\ket{\phi_2} + \ket{R}}{\sqrt{2}}) |^2  \nonumber \\=&  \frac{1}{4}|\bra{\phi_1}\h U_i^{\dd} \h U_j \ket{\phi_2} +  \bra{R}\h U_i^\dd \h U_j  \ket{R} |^2,\\
F_2=&|\bra{\phi_1}\h U_i^{\dd} \h U_j \ket{\phi_2}|^2.
\end{align}

\subsubsection{Computation of: $(H_{\psi})_{ij}=\bra{\phi_0} \hat{U}^{\dd}_i \hat{\s}_a \hat{U}^{\dd}_j\hat{H}\hat{U}_k \hat{\s}_b\hat{U}_l\ket{\phi_0}$}

The shape of $(\bm{H}_{\psi})_{ij}$ is very similar to the  matrix elements $(\bm{S}_{\psi})_{ij}$, but care needs to be taken for the Pauli string in $\h H$, which may not conserve the symmetry $\h N$. Due to the presence of up to 4 Pauli operators, which connect adjacent symmetry sectors, $\ket{R}$ needs to be chosen such that if $\expval{\h N}{\phi_0}>4$, we choose $\ket{R}=\ket{0} \otimes \ket{0}...\otimes\ket{0}$ and if $\expval{\h N}{\phi_0}\leq 3$, we can choose $\ket{R}=\ket{1} \otimes \ket{1}...\otimes\ket{1}$. We note that it is only valid if the total number of sites is strictly higher than 4.

Finally, condition III depends on the construction of the reference state, and is explained in Sec. \ref{sec:HF_prepa}.

\section{Preparation of the GHZ state with the Hartree-Fock state}
\label{sec:HF_prepa}

As part of condition III, the state $\ket{R} + \ket{\phi_0}$ needs to be implemented, where $\ket{\phi_0}$  is the minimal energy state of $H_0$ with  $n_\ua$ spin-up and $n_\da$ spin-down particles (see Eq. (34) in the main text).
To construct this superposition, we use a GHZ-state-preparation circuit as suggested in Ref. \cite{cortes2021quantum}. It requires applying a Hadamard gate on the qubit 1 (resp. $N$, where $N$ is the number of sites) and a number $(n_\ua-1)$ (resp. $n_\da-1$) of CNOT gates between qubits $i$ and $i+1$ with $i \in [1,n_\ua]$ (resp. $i \in [N+1,N+n_\da]$). Finally, we apply the $\hat{C}^\dd$ rotation operator in Eq. 34 in the main text to rotate the GHZ state in the local basis. Here,  we used the relation $\hat{C}^\dd \ket{0} {=}\ket{0}$.

The circuit depth in terms of CNOT layers is the sum of the preparation of the layers for the GHZ state preparation, which requires $\mathrm{max}(n_\ua,n_\da)-1$ layers of CNOT gates, and of the $C^\dd$ preparation, which requires $2N$ layers of CNOT gates as explained in Sec. IV of the main text.

\section{DMFT on the Bethe lattice}

For the non interacting solution, the hybridization is given by $\D^0(\w) {=}\frac{1}{2\pi} \Theta (4-\w^2) \sqrt{4-\w^2}$, where $\w$ is the real frequency, and we use the convention $\gamma=1$ as in the main text.

The discretized hybridization is given by $\D^{ED}(z) {=} \sum_{i=n_{\text{imp}}}^{N} \frac{|\e_{0i}|^2}{z - \e_{ii}}$, where we take the hopping matrix to connect a bath site to the impurity as schematically illustrated in Fig. 2 in the main text. The parameters $\e_{ij}$ are chosen to minimize the distance $D = \sum^{n_\mathrm{M}}_{n=0}|\D^{ED}(i\w_n)-\D(i\w_n)|^2$, where the hybridizations are evaluated on the Matsubara axis $\w_n=\frac{(2n+1)\pi}{\b}$, with $\b$ the inverse temperature. Here, we take $\beta{=}100$, which is small enough to keep the zero temperature formalism. Here the role of finite $\beta$ is only to get the hybridization on a finite grid on the imaginary axis, where the function is smoother. In particular, we check that the gap between the first excited state and the  ground state is large enough to have $\e^{\beta (E^{\text{excited}}-E_{\GS})}<0.01$ in order to ignore the population of the excited states compared to the GS. We use $n_\mathrm{M}{=}100$ in order to fit well the low energy part.
Once the Green's function (GF) is determined, we use the DMFT equation $\D^{\text{new}} {=} G$ to find a new bath hybridization, and we reiterate the procedure until the convergence of the GF.

Table \ref{tab:1} (resp. Table \ref{tab:2}) gives the parameters of the first (resp. last) iterations.

\begin{table}[h!]

\begin{tabular}{ccc}
$i$ & $e_{ii}$ & $e_{i1}$\\
2 &-1.17300 &-0.53714\\
3 & -0.37368 &0.38549\\
4 & -0.08996 & -0.21964\\
5  & 0.00000 & 0.13394\\
6  & 0.08996 & -0.21964\\
7  & 0.37368 & 0.38549\\
  8 & 1.17300 & 0.53714\\

\end{tabular}
\caption{\label{tab:1} Bath parameters of the fist DMFT iteration}

\end{table}

Finally, since we are at half-filling, the energy of the impurity site is given by $\e_{11}{=}-\frac{U}{2}$.

\begin{table}[h!]
\begin{tabular}{ccc}
$i$ &$ e_{ii}$ & $e{i1}$\\
2 & -4.78165 & -0.52895\\
3 & -3.09398  & 0.42680\\
4 & -2.17008 & -0.19147\\
5 & -0.00103  & 0.00418\\
6 & 1.61650 & -0.05916\\
7 & 2.74851  & 0.41342\\
  8& 4.63114  & -0.56915\\
\end{tabular}
\caption{\label{tab:2} Bath parameters of the last DMFT iteration}
\end{table}

\newpage
\section{Off-diagonal elements of the Green's function}

The algorithm proposed in Sec. IIB of the main manuscript cannot be directly applied to the off-diagonal elements of the greater GF, which are given by
\begin{align}
    G^>_{\a\b}(z)& = \bra{\text{GS}}c_{\a}(z - (H -E_{\text{GS}}))^{-1}c^\dd_{\b}\ket{\text{GS}}.
\end{align}
Here we present the extension of the method to be able to compute these off-diagonal elements. To this aim we first apply the method to obtain the elements of two GFs, defined as
\begin{align}
  G^{(1)}_{\a\b}(z) &= \bra{\GS} (\h c_\a+\h c_\b)(z-\h H+E_\mathrm{GS})^{-1}   (\h c^\dd_\a+\h c^\dd_\b)\ket{\GS},\\
  G^{(2)}_{\a\b}(z) &= \bra{\GS} (\h c_\a-i\h c_\b)(z-\h H+E_\mathrm{GS})^{-1}   (\h c^\dd_\a+i\h c^\dd_\b)\ket{\GS}.
\end{align}
With these modified GFs we then obtain
\begin{align}
  G^>_{\a\b} +G^>_{\b\a} &= G^{(1)}_{\a\b}(z) - G^>_{\a\a}(z)- G^>_{\b\b}(z) \\
  G^>_{\a\b} -G^>_{\b\a} &= i(  G^>_{\a\a}(z)+ G^>_{\b\b}(z) - G^{(2)}_{\a\b}(z))
\end{align}

So $G_{\a\b}^>$ is given by
\begin{equation}
  G_{\a\b}^> = \frac{1}{2}( G^{(1)}_{\a\b}(z) - i G^{(2)}_{\a\b}(z) + (i-1) (G^>_{\a\a}(z)+ G^>_{\b\b}(z)))
\end{equation}

The lesser GF can be computed using the same method.

\bibliography{bibli}